\documentclass[%
 reprint,
 amsmath,amssymb,
 aps,
]{revtex4-2}

\usepackage{graphicx}
\usepackage{dcolumn}
\usepackage{bm}
\usepackage{color}
\usepackage{ulem}
\DeclareMathOperator{\sech}{\textrm{sech}}

\begin{document}

\preprint{APS/123-QED}

\title{Quantum sensor network metrology with bright solitons}

\author{Dmitriy Tsarev$^{1,5}$, Stepan Osipov$^1$, Ray-Kuang Lee$^{2,3}$, Sergey Kulik$^{4,5}$ and Alexander Alodjants$^{1,5}$}
\email{alexander\_ap@list.ru}
\affiliation{$^1$ ITMO University, St. Petersburg, 197101, Russia}
\affiliation{$^{2}$ National Tsing Hua University, Hsinchu 30013, Taiwan}
\affiliation{$^{3}$ Center for Quantum Technology, Hsinchu 30013, Taiwan}
\affiliation{$^{4}$ M.V. Lomonosov Moscow State University, 1 Leninskie Gory Street, Moscow 119991, Russia}
\affiliation{$^{5}$ South Ural State University (SUSU), 76, Lenin Av., Chelyabinsk, Russia}

\begin{abstract}
We consider multiparameter quantum metrology problem with bright soliton networks in the presence of weak losses. We introduce General Heisenberg Limit (GHL) $\sigma_{\boldsymbol{\chi}}=1/N^k$ that characterizes fundamental limitations for unknown parameter measurement and estimation accuracy $\sigma_{\boldsymbol{\chi}}$ within linear ($k=1$) and nonlinear ($k=3$) quantum metrology approaches to solitons. We examine multipartite $N00N$ states specially prepared for the improvement of multiparameter estimation protocols. As a particular example of producing such states, we propose the three-mode soliton Josephson junction (TMSJJ) system as a three mode extension for the soliton Josephson junction (SJJ) bosonic model, which we previously proposed. The energy spectrum of the TMSJJ exhibits sharp phase transition peculiarities for the TMSJJ ground state. The transition occurs from a Gaussian-like (coherent) state to the superposition of entangled Fock states, which rapidly approach the three-mode $N00N$ state. We show that in the presence of weak losses, the TMSJJ enables saturate scaling relevant to the optimal state limit close to the GHL. Our findings open new prospects for quantum network sensorics with atomtronic circuits. 
\end{abstract}

\keywords{Bose-Einstein Condensate, solitons, quantum metrology, $N00N$-state}

\maketitle

\section{\label{sec:level1}Introduction}

Quantum metrology and sensorics represent a meaningful practical result of current quantum technologies~\cite{PezzeRMP, DegenRMP}. Real-world quantum metrological applications may be found in fundamental science achievements, navigation, and space, geology, life science, ecology and environment, and civil engineering, see e.g.~\cite{Crawford, Bongs}. From the practical point of view advanced quantum metrology devices and sensors require interface with networks, which may be inherent to quantum Internet in the near future, cf.~\cite{Wehner}. Thus, an urgent current goal is to study the capabilities and fundamental limitations for the measurement and estimation accuracy of distributed quantum sensors.

Typically, high-precision quantum metrology devices operate with atomic Bose-Einstein condensates (BEC)~\cite{Becker} or photonic setups~\cite{Polino}. Measurement and estimation of some unknown phase-dependent parameters inherent to atomic or photonic systems are primary in this case. It is instructive to mention linear and nonlinear quantum metrology approaches that we examine in this work. 

In the framework of linear quantum metrology, estimated phase $\phi$ linearly depends on average particle number $N$, i.e. $\phi=\chi N$, where $\chi$ is some unknown parameter that we wish to specify. In nonlinear metrology we deal with unknown nonlinear phase shift $\phi=\chi N^k$, where $k=2,3,...$. In both cases one can introduce generalized Heisenberg limit (GHL)
\begin{equation}\label{xi1}
\sigma_{GHL}^{(k)}\geq\frac{1}{N^k},
\end{equation}
that establishes fundamental ultimate accuracy $\sigma_{GHL}^{(k)}$ of one, $\chi$, parameter measurement and estimation. 

Thus, familiar linear quantum metrology operates within the Heisenberg limit (HL) obtained from~\eqref{xi1} at $k=1$. Noteworthy, the HL may be saturated by various measurement and/or detection procedures. For example,  a two-mode  Mach-Zehnder interferometer fed by the ideal $N00N$ state allows a two-mode measurement procedure saturating the HL, see e.g.~\cite{Dowling}. The detection procedure to achieve the HL may be realized in the framework of parity-measurement detection schemes~\cite{Gerry2021, Tsarev2018}. On the other hand, as we showed in~\cite{Ngo2021}, it is possible to establish a positive operator-valued measurement (POVM) procedure that enables to saturate the HL; in general, an $n$-level quantum system provides at least $n^2$ POVM elements. Theoretical studies of POVM peculiarities in high-dimension systems have been performed in a numerous number of works~\cite{Grassl,Planat,Yoshida,Pinto,Zhu, Renes}. Quantum measurements established by symmetric-informationally-complete (SIC) POVMs are optimal for quantum state tomography and were proposed in systems of various dimensions, cf.~\cite{Czerwinski, Ferrie, Brida, Lundeen}. SIC POVMs were verified experimentally for photonic low-dimension schemes including photonic circuits and spontaneous down conversion processes, see e.g.~\cite{Leuchs, Tabia, Padua, Steinberg, Ling}. These schemes are described by discrete variables; however, in this work we consider a mesoscopic number of particles that requires a continuous variable approach. In this limit, SIC POVM methods represent a great interest and are applicable, at least in theory, cf.~\cite{Ngo2021}. However, the experimental verification of SIC POVMs operating in a high-dimension photonic and/or atomic system currently looks quite cumbersome. Here, we analyse the $N00N$ state formation for mesoscopic systems described by continuous variables.

From a practical point of view, Eq.~\eqref{xi1} implies phase super-resolution that we can achieve within the $N$-particle interference. In this sense, Eq.~\eqref{xi1} helps recognizing phase super-sensitivity that may be verified by parameter (cf.~\cite{Okamoto}) 
\begin{equation}\label{S}
\mathcal{S}=\frac{1}{\sqrt{\nu N} \sigma_{\chi}},
\end{equation}
where $\nu$ is the number of trials (measurements), further we set it equal to one for simplicity; $\sigma_{\chi}$ represents the accuracy attainable for the $\chi$ parameter measurement and estimation. Establishing quantum Cramer-Rao (QCR) bound for $\sigma_{\chi}$ from~\eqref{S} we obtain 
\begin{equation}\label{S2}
 0\leq\mathcal{S}\leq\sqrt{\frac{F}{N}},
\end{equation}
where $F$ is the Fisher information related to the $\chi$ parameter measurement and estimation. Notably, if we apply inequality~\eqref{xi1} to~\eqref{S} and~\eqref{S2}, we can obtain
\begin{equation}\label{S3}
0\leq\mathcal{S}\leq N^{k-1/2}.
\end{equation}

The right part in Eq.~\eqref{S3} establishes the upper bound for phase resolution performed by a quantum sensor. In classical domain $\mathcal{S}$ obeys inequalities
\begin{equation}\label{S4}
0\leq\mathcal{S}_{cl}\leq 1,
\end{equation}
that may be achieved in the framework of the linear metrology ($k=1$) approach with coherent (Glauber's) states. As it follows from~\eqref{S2}-\eqref{S4}, purely quantum sensitivity for $\mathcal{S}$, that is $\mathcal{S}_q\equiv\mathcal{S}>1$, requires achievement of quantum Fisher information $F$ beyond value $F\simeq \sqrt{N}$, which is relevant to the standard quantum limit (SQL) of phase estimation.

At $k>1$ Eq.~\eqref{xi1} defines the super-Heisenberg limit (SHL) that enables to determine the ultimate accuracy of unknown parameter measurement and estimation within the nonlinear quantum metrology. In this case, the HL can be overcome even with Glauber's coherent states due to nonlinearity~\cite{Maldonado}. It is shown that two-mode $N00N$ state can saturate the SHL with $k=2$ within an unknown nonlinear phase shift estimation procedure. The SHL implies the use of squeezing and nonlinear properties of a material (atomic or photonic) system, cf.~\cite{Napolitano}. 

Previously, in~\cite{Tsarev2019}, we showed that quantum bright solitons provide the maximal value of degree $k=3$ that may be obtained with Kerr-like medium due to soliton spatial degrees of freedom. At the same time, we proposed the soliton Josephson junction (SJJ) device, which enables to produce Fock state superposition close to $N00N$ states and protected against small number of particle losses~\cite{Alodjants2022}. It is important to stress that $N00N$ as well as ``super-entangled" states may be achieved with weakly-attracting particles, which corresponds to negative scattering length, cf.~\cite{Mazzarella,Drummond}. 

\begin{figure*}[t]
\begin{minipage}[h]{0.69\linewidth}
\center{\includegraphics[width=0.8\linewidth]{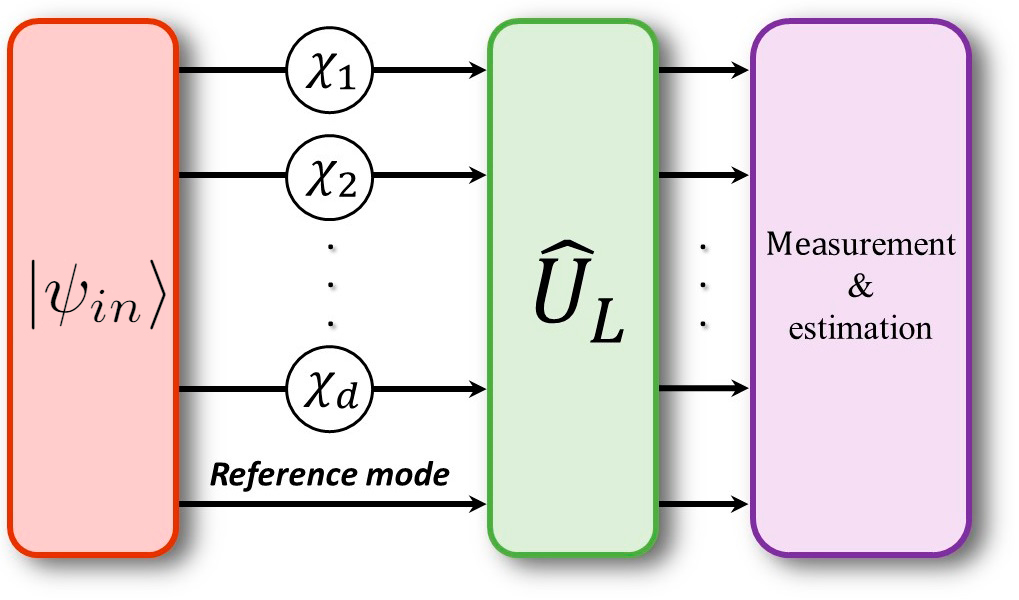}}\\ a) 
\end{minipage}
\hfill
\begin{minipage}[h]{0.3\linewidth}
\center{\includegraphics[width=1\linewidth]{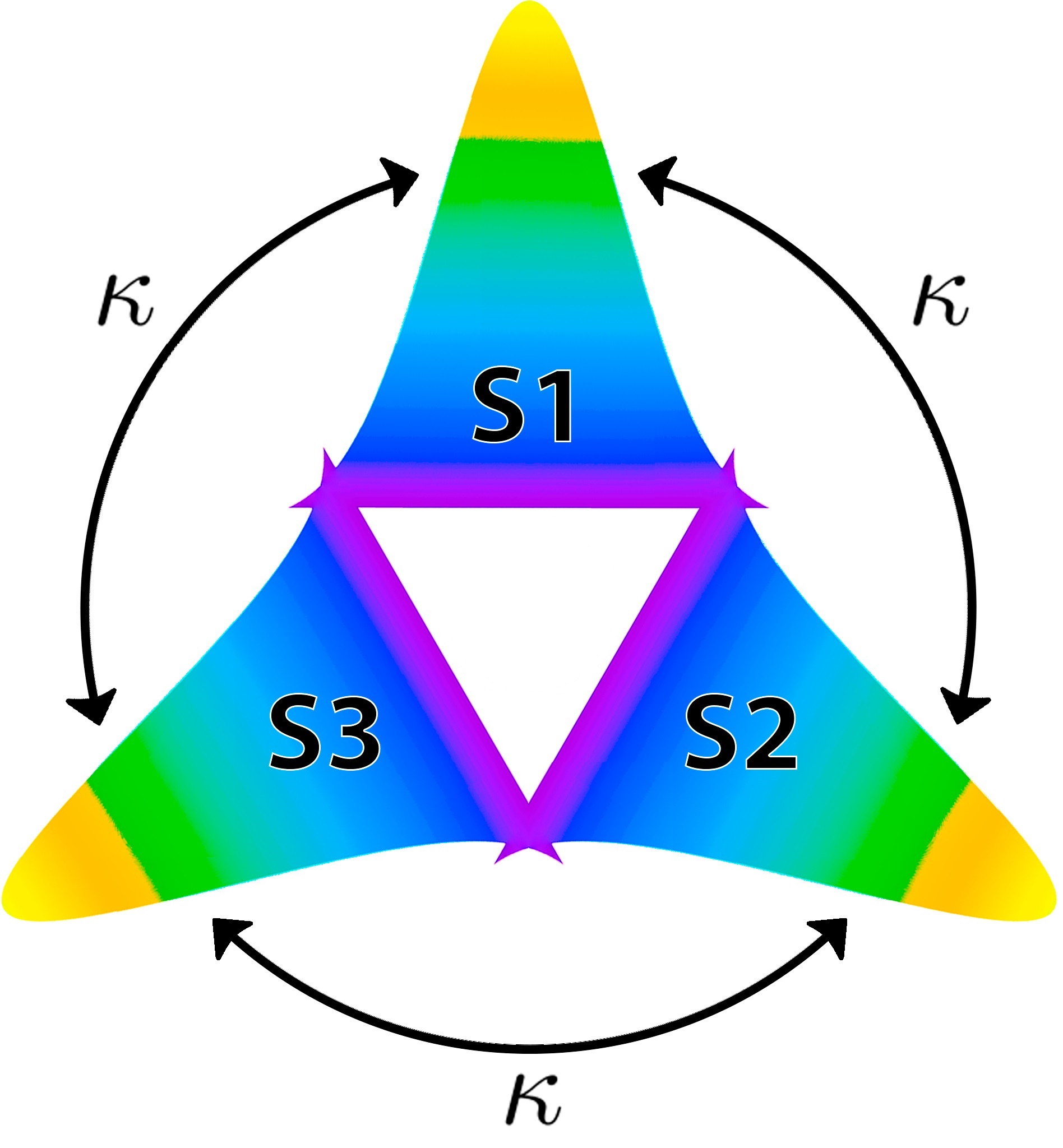}} \\b)
\end{minipage}
\caption{(a) Sketch of multiparameter quantum metrology circuit with solitons. $\left|\psi_{in}\right\rangle$ is a multipartite state of quantum solitons prepared for metrological tasks. This state distributes within the QSN and accumulates phases $\phi_j$ containing information about estimated parameters $\chi_j$ ($j=1,...,d$). Operator $\hat{U}_L$ denotes the action of network BS that allows to build measurement procedure of unknown parameters with their estimation. Other details are given in the text. (b) Solitons layout for balanced tripartite $N00N$ state preparation. The solitons are trapped in a three-well potential (not shown) providing each-to-each tunnel coupling. The symmetry of the system tells invariance under cyclic permutation of phase and particle difference variables, respectively. The double-sided arrows illustrate tunnel couplings between the solitons.}
\label{FIG:Scheme}
\end{figure*}

Quantum networks bring new advantages and opportunities to quantum sensorics~\cite{Zheshen,Goldberg}. In practise, on-chip quantum sensor networks (QSN) may be implemented by atomtronic~\cite{Amico} or photonic~\cite{Flamini, Clements,Fldzhyan} circuit facilities. Especially we would like to mention here CMOS compatible platforms~\cite{Moss}; in particular, microrings possessing Kerr-like nonlinearity are capable for photonic soliton lattice formation~\cite{Karpov} and all-to-all entanglement achievement~\cite{Lukin}.

Recently, fundamental aspects of multiparameter sensorics and metrology have become the subject of intensive study~\cite{Jing,Gessner}. Various measurement strategies and procedures are discussed within simple two-mode phase estimation schemes~\cite{Pezze,Bringewatt}. Capacity of non-classical states aimed at improvement of overall metrological accuracy achieved within QSNs represents a primary task that has not fully studied yet, see e.g.~\cite{Humphreys,Hong}. 

In this work, we continue our studies on quantum metrology with solitons, established within the two-mode approach~\cite{Tsarev2018, Tsarev2019, Ngo2021, Tsarev2020, Alodjants2022, Ngo}. In~\cite{Tsarev2018, Tsarev2019, Tsarev2020, Alodjants2022} we discussed in detail atomic BECs possessing negative scattering length as a physical platform for metrology and SJJ realization in practice, cf.~\cite{Khaykovich2002}. The influence of losses and decoherence was analysed in~\cite{Tsarev2020, Alodjants2022, Ngo}. Notice, our current proposal may be also realized in quantum optics with coupled optical fibers/waveguides possessing the Kerr-like nonlinearity, see e.g.~\cite{Spagnolo}.

The paper is arranged as follows. In Sec. II, we analyse the fundamental limits for multimode (multiparameter) nonlinear quantum metrology with quantum solitons spatially distributed within some QSN and established in Fig.~\ref{FIG:Scheme}a. We specify some peculiarities for parameter accuracy estimation resulting from the implementation of the spatially distributed multipartite $N00N$ state. Then, we examine the multiparameter metrology and sensing task in the practically important two-parameter quantum metrology limit. In Sec. III, we give a general description for a novel three-mode soliton Josephson junction (TMSJJ) model for metrological applications. We show how to obtain a three-mode $N00N$-like (entangled Fock) state by coupled bright solitons containing a mesoscopic number of particles. First, We discuss a semiclassical TMSJJ model for atomic BECs trapped in a symmetric three-well potential. The geometry of the TMSJJ is presented in Fig.~\ref{FIG:Scheme}b. To be more specific, we analyse a completely symmetric case of soliton couplings, cf.~\cite{Compagno,Nemoto,Guo}. Second, for the full-quantum TMSJJ model we examine the energy spectrum exhibiting a phase transition to entangled Fock state that anticipates three-mode $N00N$ state formation. In Sec. IV, we combine these results accounting losses that eventually occur in the metrological scheme during quantum state evolution, see Fig.~\ref{FIG:Scheme}a. We examine a complete three-mode soliton metrology task that includes the three-mode $N00N$ state preparation, phase accumulation, and measurement procedure. The multiparameter estimation bounds with quantum solitons in the presence of losses are elucidated using the upper bound of Fisher information. We show that in the framework of linear and nonlinear metrologies the TMSJJ allows approaching the GHL even with weak particle losses. In Conclusion we summarize the results obtained.

\section{\label{sec:level1} Fundamental limits of multiparameter nonlinear quantum metrology}

Consider the measurement and estimation procedure for a set of unknown parameters $\chi_j$, shown in Fig.~\ref{FIG:Scheme}a, and exploiting $(n=d+1)$-partite ($(d+1)$-mode) spatially entangled $N00N$ state that we establish as 
\begin{eqnarray}\label{psi_in}
\nonumber 
\left|\psi_{in}\right\rangle &=& \varepsilon\big(\left|0,N,0,...,0\right\rangle + \left|0,0,N,...,0\right\rangle + ... \\ 
 &+& \left|0,0,0,...,N\right\rangle\big) + \sqrt{1 - \varepsilon^2 d}\left| N,0,0,...0\right\rangle,
\end{eqnarray}
where $\varepsilon\neq0$ describes the amplitude of the  estimated channels in Fig.~\ref{FIG:Scheme}a. Then, we assume that the $\left|\psi_{in}\right\rangle$ state is distributed over the QSN nodes accumulating unknown phase shifts $\phi_j=\chi_j N^k$, $j=1,...,d$. Thus, after transforming $\left|\psi_{in}\right\rangle$, we obtain
\begin{eqnarray}\label{psi_n}
\nonumber
&&\left|\psi_n\right\rangle = \varepsilon\big(e^{i\phi_1}\left|0,N,0,...,0\right\rangle + e^{i\phi_2}\left|0,0,N,...,0\right\rangle + ... \\ 
 && + e^{i\phi_d}\left|0,0,0,...,N\right\rangle\big) + \sqrt{1 - \varepsilon^2 d}\left| N,0,0,...,0\right\rangle.
\end{eqnarray}

The QSN capacity corresponds to state $\left|\psi_n\right\rangle$ and consists of simultaneous estimation up to $d$ phase parameters $\chi_j$ in respect of the reference mode (the last term in~\eqref{psi_n}). 

Eq.~\eqref{psi_n} with $k = 1$ corresponds to the linear metrology approach, while $k > 1$, $k\in \boldsymbol N$, establishes the nonlinear quantum metrology limit. Particularly, $k=2$, if we use routine Kerr-like mediums for unknown phase shifts in Fig.~\ref{FIG:Scheme}a and plane waves description, cf.~\cite{Maldonado}. For the nonlinear quantum metrology with a soliton network we can take $k=3$, cf.~\cite{Alodjants2022}. 

In the framework of multiparameter quantum metrology, we are interested in minimizing the overall variance 
\begin{equation}\label{over}
\sigma_{\boldsymbol{\chi}}\equiv \Big( \sum_{i=1}^d\sigma^2_{\chi_j}\Big)^{1/2}, 
\end{equation}
where $\boldsymbol{\chi}\equiv\{\chi_j\}$ denotes a set of unknown parameters, and $\sigma_{\chi_j}$ is an accuracy of their simultaneous measurement and estimation that we characterize by quantum Fisher information (QFI). In a general case of multiparameter estimation, the QFI, $\hat{F}$, represents a $d\times d$ matrix, where $d$ is the number of the parameters to be simultaneously estimated, cf.~\cite{Gessner}. The QFI matrix elements take a form
\begin{eqnarray}\label{QFI_d}
\nonumber
 &&F_{ij} = 4\textrm{Re}\Big[\big\langle\partial_{\chi_i}\psi_n\big|\partial_{\chi_j}\psi_n\big\rangle \\ 
 &&\;\;\;\;\;\;\;\;\;\;\;\;\;\;\;\;\;\;\;\;\;\;\;\;\;\;\;\; - \big\langle\partial_{\chi_i}\psi_n\big|\psi_n\big\rangle\big\langle\psi_n\big|\partial_{\chi_j}\psi_n\big\rangle\Big],
\end{eqnarray}
where $\left|\psi_n\right\rangle$ is some $n$-mode state with $n\geq d+1$; $\chi_{i,j}$ are measurables, some phase parameters depending on $N$; $\left|\partial_{\chi_{i,j}}\psi_n\right\rangle\equiv\frac{\partial}{\partial\chi_{i,j}}\left|\psi_n\right\rangle$. The measurement (overall) accuracy is limited by the QCR bound,
which for $\hat{F} = \{F_{ij}\}$ is
\begin{equation}\label{QCR_d1}
\sigma_{\boldsymbol{\chi}}\geq\Big(\textrm{Tr}\left(\hat{F}^{-1}\right)\Big)^{1/2}.
\end{equation}

Substituting~\eqref{psi_n} into~\eqref{QFI_d} we obtain
\begin{equation}\label{QFI}
 F_{ij} = 4N^{2k}\varepsilon^2(\delta_{ij}-\varepsilon^2),
\end{equation}
which gives $\textrm{Tr}(\hat{F}^{-1}) = \frac{1}{N^{2k}}\frac{d(1 + \varepsilon^2 - d\varepsilon^2)}{4\varepsilon^2}(1 - d\varepsilon^2)$.

Thus, for the balanced $N00N$ state with $\varepsilon=1/\sqrt{d+1}$ we obtain
\begin{equation}\label{QFI3}
\sigma_{\boldsymbol{\chi}}\geq\frac{1}{N^k}\sqrt{\frac{d(d+1)}{2}}.
\end{equation}
In particular, for the two-mode $N00N$ state metrology we must take $d=1$, and~\eqref{QFI3} leads to the GHL established in~\eqref{xi1}. 

At $d>1$ the overall accuracy $\sigma_{\boldsymbol{\chi}}$ degrades, and one can obtain $\sigma_{\boldsymbol{\chi}}>\sigma_{GHL}^{(k)}$. For the three-mode $N00N$ state metrology, that we examine below, $d=2$, the ultimate precision is $\sqrt{3}\sigma_{GHL}^{(k)}$, obtained for the balanced $N00N$ state at $\varepsilon=1/\sqrt{3}$, cf.~\eqref{psi_n}. This limit we can overcome with non-balanced $N00N$ state setting $\varepsilon = 1/\sqrt{d+\sqrt{d}}$ in~\eqref{psi_n},  cf.~\cite{Humphreys}:
\begin{equation}\label{QFI3_2}
\sigma_{\boldsymbol{\chi}}\geq\frac{1}
{N^k}\frac{\sqrt{d} (\sqrt{d}+1)}{2},
\end{equation}
which for $d=2$ approaches $\sigma_{\boldsymbol{\chi}}\geq (1+1/\sqrt{2})\sigma_{GHL}^{(k)}\simeq \sqrt{2.914}\sigma_{GHL}^{(k)}$, that gives a small advantage in comparison with balanced $N00N$ state, and the preparation of such optimized states is even more complicated. Further, we refer $\sigma_{OS}^{(k)} =  \sqrt{2.914}\sigma_{GHL}^{(k)}$ as the $N00N$ optimized state (OS) limit, while the main focus is made on the $N00N$-state-based metrology.

\section{\label{sec:level1} TMSJJ model for tripartite $N00N$ state preparation}

\subsection{\label{sec:level1} Background}

The preparation of state~\eqref{psi_in} and/or~\eqref{psi_n} for arbitrary large $d$ represents a nontrivial practical task. In this work, we examine a realistic situation of quantum metrology with the TMSJJ ($n=3$) that enables to prepare $\left|\psi_{in}\right\rangle$ or $\left|\psi_{n}\right\rangle$ close to the tripartite $N00N$ state, cf.~\eqref{psi_in},~\eqref{psi_n}.

In optics, quantum state $\left|\psi_{in}\right\rangle$, which characterizes bright solitons possessing a mesoscopic photon number (up to few hundreds), may be created in semiconductor microstuctures with strong nonlinearities~\cite{Sich}. Then, such solitons can be used for propagation in linear waveguide circuits, which allow operating with unknown phase parameters, see Fig.~\ref{FIG:Scheme}b, cf.~\cite{Skryabin}. On the other hand, microcavities, based on low-loss $Si_3N_4$ microrings, may be exploited for soliton lattice formation~\cite{Karpov}. In this case, weak coupling between solitons originates from quantum superposition and vanishing overlapping of solitons in the ring, cf. \cite{Ngo2021}. Alternatively, we can arrange photonic molecules by microring weak coupling and choosing an appropriate free spectral range for them, cf.~\cite{Liao}.

Below we examine another possibility to realize quantum metrology and sensing with atomic TMSJJ, which is shown in Fig.~\ref{FIG:Scheme}b and provides the preparation of states $\left|\psi_{in}\right\rangle$ and $\left|\psi_{n}\right\rangle$, respectively,  cf.~\cite{Alodjants2022}.

\subsection{\label{sec:level1} Semiclassical TMSJJ model}

Consider the Hartree (variational) approach to the TMSJJ that represents generalization of the two-component SJJ, cf.~\cite{Tsarev2019, Tsarev2018}. In Fig.~\ref{FIG:Scheme}b we establish the geometry of arranged solitons. The Hamiltonian in the second quantisation form may be written as 
(cf.~\cite{Tsarev2018, Ngo2021, Raghavan2000})
\begin{equation}\label{Ham1}
\hat{H} = \sum\limits_{j=1}^3\hat{a}_j^\dag\left(-\frac{1}{2}\frac{\partial^2}{\partial x^2} - \frac{u}{2}\hat{a}_j^\dag\hat{a}_j\right)\hat{a}_j - \kappa\sum\limits_{j=1}^3\sum_{i\neq j}\hat{a}^\dag_{i}\hat{a}_j,
\end{equation}
where $\hat{a}_j\equiv \hat{a}_j(x)$ is the bosonic annihilation operator obeying commutation rule $[\hat{a}_i(x),\hat{a}_j^\dag(x')] = \delta_{i,j}\delta(x-x')$. In particular, in atomtronics we can assume that condensates are placed within three symmetrically arranged cigar-shaped each-to-each coupled traps, as shown in Fig.~\ref{FIG:Scheme}b. The nonlinear particle interaction parameter, $u = 2\pi|a_{sc}|/r_0$, is responsible for Kerr-like nonlinearity~\cite{Raghavan2000}; $r_0 = \sqrt{\hbar/M\omega_0}$ is the characteristic trap scale in the transverse direction; $M$ is the particle mass; $\omega_0$ is the characteristic harmonic trap frequency; $a_{sc}$ is the BEC particle scattering length. For bright matter solitons we consider BEC of attractive particles, such as $^{7}$Li, for which $a_{sc}<0$. We take tunneling coupling constant $\kappa$ the same for all coupling links between the solitons. The variational state for the system in Fig.~\ref{FIG:Scheme}b we chose as (\cite{Lai1989,Lai1989a,Alodjants1995}) 
\begin{widetext}
\begin{eqnarray}\label{vector}
\left|\Psi_N\right\rangle = \frac{1}{\sqrt{N!}}\Bigg[\int_{-\infty}^\infty\Big(\psi_1(x)\hat{a}_1^\dag(x)+\psi_2(x)\hat{a}_2^\dag(x)+\psi_3(x)\hat{a}_3^\dag(x)\Big)dx\Bigg]^N\left|0\right\rangle,
\end{eqnarray}
\end{widetext}
where $|0\rangle\equiv|0_1,0_2,0_3\rangle$ denotes the three-mode vacuum state; $N$ is the total number of particles; $\psi_j(x)$ ($j=1,2,3$) is the unknown variational function obeying the normalization condition
\begin{equation}\label{norm}
\sum_{j=1}^3\int_{-\infty}^\infty \left|\psi_j(x)\right|^2dx = \sum_{j=1}^3\frac{N_j}{N}~\equiv \sum_{j=1}^3 n_j = 1,
\end{equation} 
where $0\leq n_j\leq 1$ is the fraction of particles populating the $j$-th well. Bosonic creation and annihilation operators act on total state $\left|\Psi_N\right\rangle$ in~\eqref{vector} as following:
\begin{eqnarray}\label{action}
&&\hat{a}_j^\dag(x)\left|\Psi_N\right\rangle = \sqrt{N+1}\psi_j^*(x)\left|\Psi_{N+1}\right\rangle;\nonumber\\
&&
\hat{a}_j(x)\left|\Psi_N\right\rangle = \sqrt{N}\psi_j(x)\left|\Psi_{N-1}\right\rangle.
\end{eqnarray}
The Hamilton function in the Hartree approximation may be obtained from~\eqref{Ham1} with~\eqref{vector},~\eqref{action} and reads as
\begin{widetext}
\begin{eqnarray}\label{Ham2}
H = \langle\Psi_N|\hat{H}|\Psi_N\rangle = N\sum_j\Bigg(\frac{1}{2}\left|\frac{\partial\psi_j}{\partial x}\right|^2- \frac{u(N-1)}{2}\left|\psi_j\right|^4 - \kappa\sum_{i\neq j}\psi_i^*\psi_j\Bigg).
\end{eqnarray}
\end{widetext}
Eq.~\eqref{Ham2} implies coupled Gross-Pitaevskii equations 
\begin{eqnarray}\label{Schr_eq}
&&i\dot\psi_j=-\frac{1}{2}\frac{\partial^2}{\partial x^2}\psi_j - u(N-1)\left |\psi_j\right|^2\psi_j - \kappa\psi_m - \kappa\psi_k,\nonumber\\
&&
j, m, k=1,2,3, \quad m\neq j\neq k.
\end{eqnarray}

In the limit of the absence of coupling, i.e. at $\kappa=0$ (condensates are isolated within their traps), Eqs.~\eqref{Schr_eq} possess separable bright soliton solutions, which look like
\begin{equation}\label{ansatz}
\psi_j=n_j\frac{\sqrt{u(N-1)}}{2}\sech\left[\frac{u(N-1)}{2}n_jx\right]e^{i\theta_j},
\end{equation}
where $\theta_j=\frac{u^2(N-1)^2n_j^2}{8}t$ is the $j$-th soliton phase; $j = 1, 2, 3$. 

The variational approach presumes that soliton populations $n_j$ and phases $\theta_j$ become time-dependent if weak coupling between the solitons is realised, $\kappa\neq0$. Substituting~\eqref{ansatz} into~\eqref{Ham2} and integrating over the space variable we obtain 
\begin{eqnarray}\label{Ham3}
H_{eff}&=&\frac{1}{N}\int_{-\infty}^\infty{Hdx} \\
&=&-2\kappa\sum_j\left(\frac{\Lambda}{3}n_j^3 + \frac{1}{4}\sum_{i\neq j}I_{ij}\cos\left[\theta_j-\theta_i\right]\right), \nonumber
\end{eqnarray}
where $I_{ij}~\equiv n_{ij}\left(1-z_{ij}^2\right)\left(1 - 0.21z_{ij}^2\right)$; $n_{ij} = n_j + n_i$; $z_{ij}=(n_j-n_i)/n_{ij}$ is the population imbalance between the $i$-th and $j$-th solitons; $\Lambda = \frac{u^2(N-1)^2}{16\kappa}$ is the vital parameter that governs TMSJJ various dynamical regimes. Notice,~\eqref{Ham3} describes the energy of the system per particle.

Eq.~\eqref{Ham3} establishes the TMSJJ model in the Hartree approximation possessing two mutually conjugated sets of variables $\{n_j\}$ and $\{\theta_j\}$. From equations $\frac{\partial n_j}{\partial t} = \frac{\partial H_{eff}}{\partial\theta_j}$ and $\frac{\partial\theta_j}{\partial t} = - \frac{\partial H_{eff}}{\partial n_j}$ we obtain 
\begin{widetext}
\begin{subequations}\label{master_eqs}
\begin{eqnarray}
\dot{n}_1&=&\frac{n_{31}}{2}\left(1-z_{31}^2\right)\left(1-0.21z_{31}^2\right)\sin\left[\Theta_{31}\right] - \frac{n_{12}}{2}\left(1-z_{12}^2\right)\left(1-0.21z_{12}^2\right)\sin\left[\Theta_{12}\right];\label{master_eq_a}\\ \nonumber
\dot{\Theta}_{12}&=&\Lambda n_{12}^2z_{12} - 2z_{12}\left[1.21-0.42z_{12}^2\right]\cos\left[\Theta_{12}\right] \\ \nonumber
&+&\Big(\frac{1}{2}\left(1 - z_{23}^2\right)\left(1-0.21 z_{23}^2\right)
+ \frac{2n_3z_{23}}{n_{23}}\left[1.21-0.42z_{23}^2\right]\Big)\cos\left[\Theta_{23}\right]\\
&-&\Big(\frac{1}{2}\left(1 - z_{31}^2\right)\left(1-0.21 z_{31}^2\right)- \frac{2n_3z_{31}}{n_{31}}\left[1.21 - 0.42z_{31}^2\right]\Big)\cos\left[\Theta_{31}\right], \label{master_eq_b} 
\end{eqnarray}
\end{subequations}
\end{widetext}
where $\Theta_{ij} = \theta_j - \theta_i$ (note that $\sum\Theta_{ij} = 0$); the dots in~\eqref{master_eqs} denote the derivatives with respect to renormalized time $\tau = 2\kappa t$. Equations for other four variables $n_2$, $n_3$ and $\Theta_{23}$, $\Theta_{31}$ can be explicitly obtained from~\eqref{master_eq_a} and~\eqref{master_eq_b} with cyclic permutation of indices $i,j=1,2,3$.

We are interested in stationary solution of Eqs.~\eqref{master_eqs} assuming $\dot{n}_j = 0$ and $\dot{\Theta}_{ij} = 0$. In general, these solutions correspond to the entangled Schr{\"o}dinger-Cat-like (SC) states, which admit $N00N$ states formation in some limit, cf.~\cite{Tsarev2018}. In this work, we restrict ourselves by examining a complete set of Eqs.~\eqref{master_eqs} useful for the $N00N$ states. 

In particular, let us suppose $n_2 = n_3 = \delta$ and $n_1 = 1 - 2\delta$ ($\delta\rightarrow0$), when all particles may be accumulated in a one soliton state. In this limit Eqs.~\eqref{master_eqs} lead to
\begin{equation}\label{phase_eq}
 \cos\left[\Theta_{12}\right] = \cos\left[\Theta_{31}\right] = \frac{\Lambda - 0.5\cos\left[\Theta_{23}\right]}{1.58}.
\end{equation}

The three-mode quantum metrology scheme that we consider below, requires one mode to be reference leaving us two modes that accumulate phase shifts in respect to the reference one. In the paper, we examine two particular cases: these phase-shifted modes are either out-of-phase or in-phase. In particular, for the out-of-phase shifts we take for~\eqref{phase_eq} $\Theta_{12} = \Theta_{31}~\equiv \Theta_-$, $\Theta_{23} = -2\Theta_-$ and obtain for soliton phase 
\begin{equation}\label{out_of_phase-n00n}
\cos\left[\Theta_-\right] = \sqrt{1.124 + \Lambda} - 0.79
\end{equation}
existing only at $\Lambda\leq2.08$. 

For the second, in-phase shifts, limit, we take $\Theta_{12} = -\Theta_{31}~\equiv \Theta_+$, $\Theta_{23} = 0$ and obtain another solution
\begin{equation}\label{in_phase-n00n}
\cos\left[\Theta_+\right] = \frac{\Lambda - 0.5}{1.58},
\end{equation}
which is also valid for $\Lambda\leq2.08$. 

Notice, at $\Lambda = 2.08$ both solutions coincide at $\Theta_\pm = 0$ providing another important special case of the $N00N$ state preparation, which we discuss below.

The stationary solutions of~\eqref{master_eqs} under consideration imply $n_1 \approx 1$ or, similarly, $N_1 \approx N$ and $N_2 \approx N_3 \approx 0$ that form state $\left|N,0,0\right\rangle$ for the first mode of the three-mode $N00N$ state. In the same manner we can find solutions for the other, $n_2 \approx 1$ and $n_3 \approx 1$, modes involved in the $N00N$ state. The explicit form of the three-mode $N00N$ state that may be obtained from~\eqref{vector} and~\eqref{ansatz} and takes into account~\eqref{phase_eq} looks like
\begin{widetext}
\begin{eqnarray}\label{TMSJJ_n00n}
|N00N\rangle_{\pm}=\frac{1}{\sqrt{3}}\big(\left|N,0,0\right\rangle +e^{iN\Theta_\pm}\left|0,N,0\right\rangle+ e^{\pm iN\Theta_\pm}\left|0,0,N\right\rangle\big),
\end{eqnarray}
\end{widetext}
where $\pm$ subscripts identify the in- and out-of-phase $N00N$ states. In~\eqref{TMSJJ_n00n} we presume that the first channel of the interferometer is a reference one, setting formally $\theta_1 = 0$. 

Thus, we can associate each of states $|N00N\rangle_{\pm}$ in~\eqref{TMSJJ_n00n} with state $|\psi_n\rangle$, see~\eqref{psi_n}, that may be used in quantum metrology to estimate the parameters embodied in phases $\Theta_{\pm}$. In this case $\Theta_{\pm}$ directly relate to unknown parameters $\chi_j$ shown in Fig.~\ref{FIG:Scheme}a.

\subsection{\label{sec:level1}Quantum TMSJJ model}

To develop a fully quantum TMSJJ model, it is necessary to quantize effective Hamiltonian~\eqref{Ham3}. The quantization procedure that we use below is similar to the prescribed in~\cite{Tsarev2020}. 

First, the number of particles populating each of the solitons we describe by operators $\hat{N}_i=\hat{a}_i^\dag\hat{a}_i$, $i=1,2,3$.

Second, we represent annihilation operators $\hat{a}_i$ as $\hat{a}_i = \sqrt{\hat N_i}e^{i\hat \theta_i}$, cf.~\cite{Paraoanu}. Thus, one can use mapping $2N\sqrt{n_in_j}\cos\left [\Theta_{ij}\right] \rightarrow \left(\hat{a}_i^\dag \hat{a}_j + \hat{a}_j^\dag\hat{a}_i\right)$, $i,j = 1,2,3$, $i\neq j$. We also introduce relative population imbalance operator $\hat{z}_{ij}=\frac{\hat{a}_j^\dag\hat{a}_j - \hat{a}_i^\dag\hat{a}_i}{\hat{a}_j^\dag\hat{a}_j + \hat{a}_i^\dag\hat{a}_i}$ and formally establish $\sqrt{1-\hat z_{ij}^2}$ in the Taylor series form as
\\
\begin{equation}
\sqrt{1-\hat z_{ij}^2}=\sum_{k=0}^\infty\left(-1\right)^k C_{0.5}^k \hat z_{ij}^{2k},
\end{equation}
where $C_{0.5}^k = \frac{1}{k!}\prod_{l=0}^{k-1}(0.5 - l)$. Finally, the TMSJJ quantum Hamiltonian in the second quantization form looks like (cf.~\eqref{Ham3})
\begin{widetext}
\begin{equation}\label{Ham6}
\hat{H}_{TMSJJ}=2\kappa\Bigg(-\frac{\Lambda}{3N^3}\sum_{i}\left(\hat{a}_i^\dag\hat{a}_i\right)^3 -\frac{1}{8N}\Bigg\{\sum_{i\neq j}\sum_{k=0}^\infty\left(-1\right)^k C_{0.5}^k\left(1-0.21 \hat{z}_{ij}^2\right)\left(\hat{a}_i^\dag\hat{a}_j + \hat{a}_j^\dag\hat{a}_i\right)\hat z_{ij}^{2k} + H.C.\Bigg\}\Bigg),
\end{equation}
where H.C. stands for the Hermitian conjugate.

We characterize tripartite quantum state of coupled solitons without losses in generally as 
\begin{equation}\label{Fock-vector}
\left|\Psi\left(\tau\right)\right\rangle = \sum_{N_1=0}^N\sum_{N_2=0}^{N-N_1} A_{N_1,N_2}\left(\tau\right)\left|N_1,N_2,N_3\right\rangle,
\end{equation}
where $N_1 + N_2 + N_3 = N = \textrm{const}$; $\tau = 2\kappa t$. Coefficients $A_{N_1,N_2}\left(\tau\right)$ in~\eqref{Fock-vector} obey Shcr{\"o}dinger equation
\begin{equation}\label{SC_eq}
i\frac{\partial}{\partial \tau}A_{N_1,N_2}\left(\tau\right) = \left\langle N_1,N_2,N_3\left|\hat H_{SJJ}\right|\Psi\left(\tau\right)\right\rangle.
\end{equation}
Substituting~\eqref{Fock-vector} and~\eqref{Ham6} into~\eqref{SC_eq} we obtain
\begin{align}\label{SC_eq2}
\nonumber
i\dot{A}_{N_1,N_2}\left(\tau\right) = \alpha_{N_1,N_2} \left( \Lambda \right) A_{N_1,N_2} &+\beta_{N_1,N_2}A_{N_1-1,N_2+1}+\beta_{N_2,N_1}A_{N_1+1,N_2-1}\\
&+\beta_{N_2,N_3}A_{N_1,N_2-1}+\beta_{N_3,N_2}A_{N_1,N_2+1}+\beta_{N_3,N_1}A_{N_1+1,N_2}+\beta_{N_1,N_3}A_{N_1-1,N_2},
\end{align}
where we made definitions
\begin{subequations}\label{SC_eq3}
\begin{flalign}
&\alpha_{N_i,N_j}=-\frac{\Lambda}{3}\frac{N_i^3+N_j^3+\left( N - N_i - N_j \right)^3}{N^3};&
\end{flalign}
\begin{align}
\nonumber
\beta_{N_i,N_j}=-\frac{1}{2N}\frac{1}{N_i+N_j}\bigg(\left(N_j + 1 \right)\sqrt{N_i\left(N_i-1 \right)}\left [ 1-0.21\left(\frac{N_j-N_i}{N_j+N_i} \right)^2\right]&\\
+N_i\sqrt{N_j\left(N_j+1\right)}&\left [ 1-0.21 \left( \frac{N_j-N_i+2}{N_j+N_i} \right)^2 \right ] \bigg).
\end{align}
\end{subequations}
\end{widetext}

\begin{figure*}[t]
\begin{minipage}[h]{0.32\linewidth}
\center{\includegraphics[width=1\linewidth]{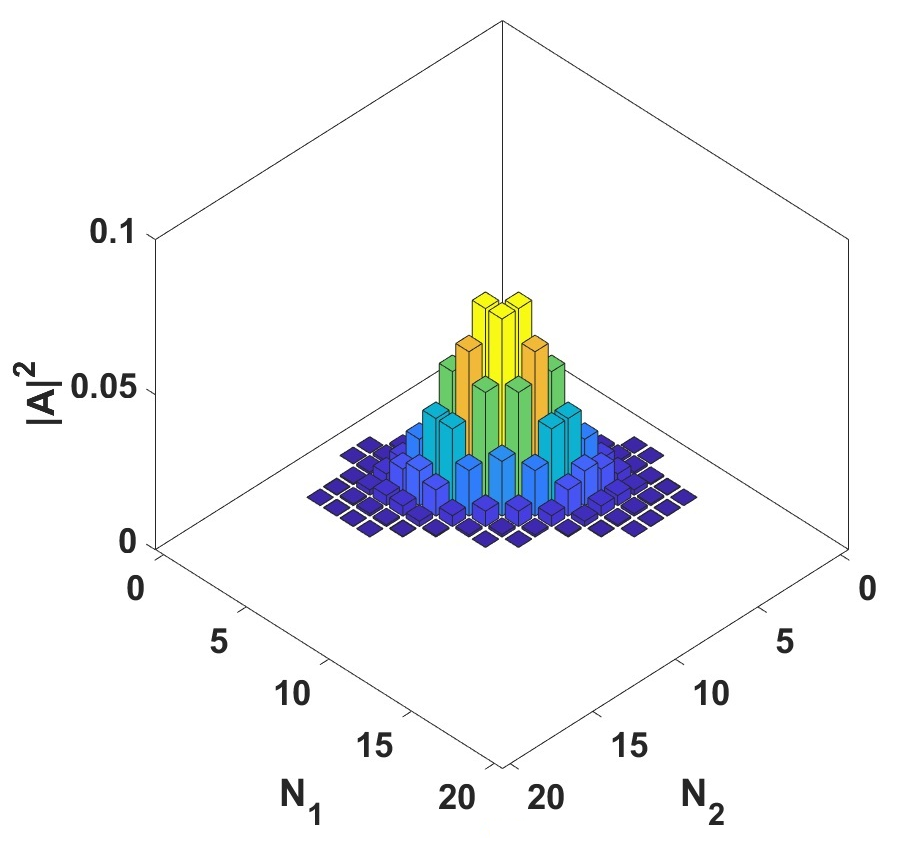}}\\ a) 
\end{minipage}
\hfill
\begin{minipage}[h]{0.32\linewidth}
\center{\includegraphics[width=1\linewidth]{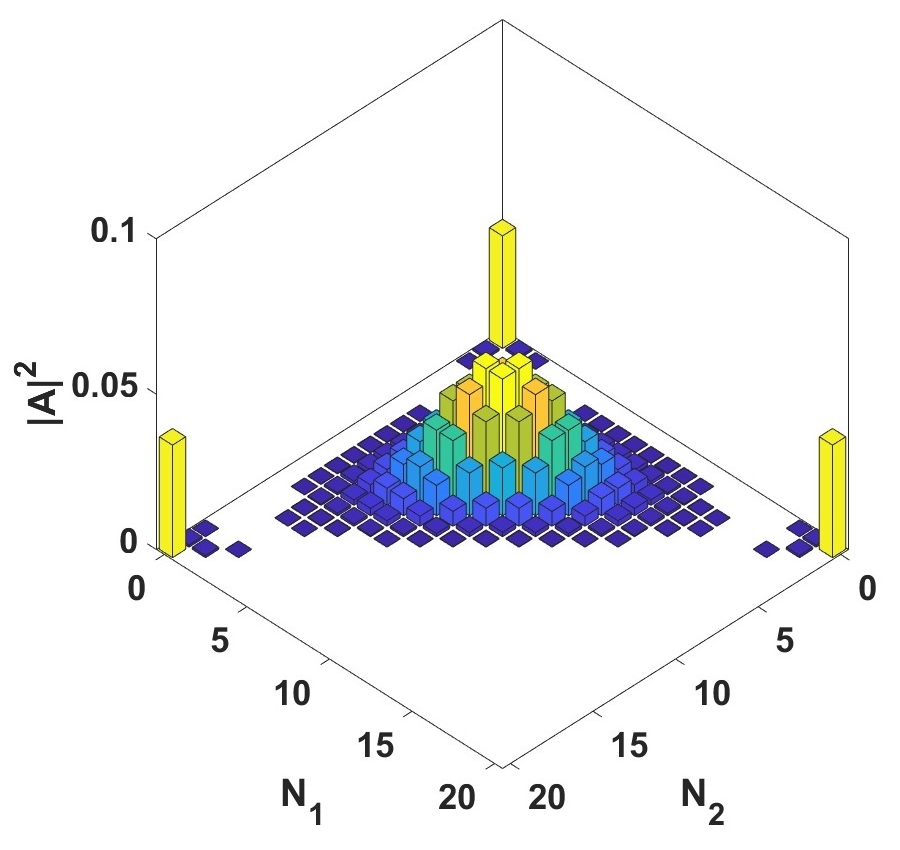}} \\b)
\end{minipage}
\hfill
\begin{minipage}[h]{0.32\linewidth}
\center{\includegraphics[width=1\linewidth]{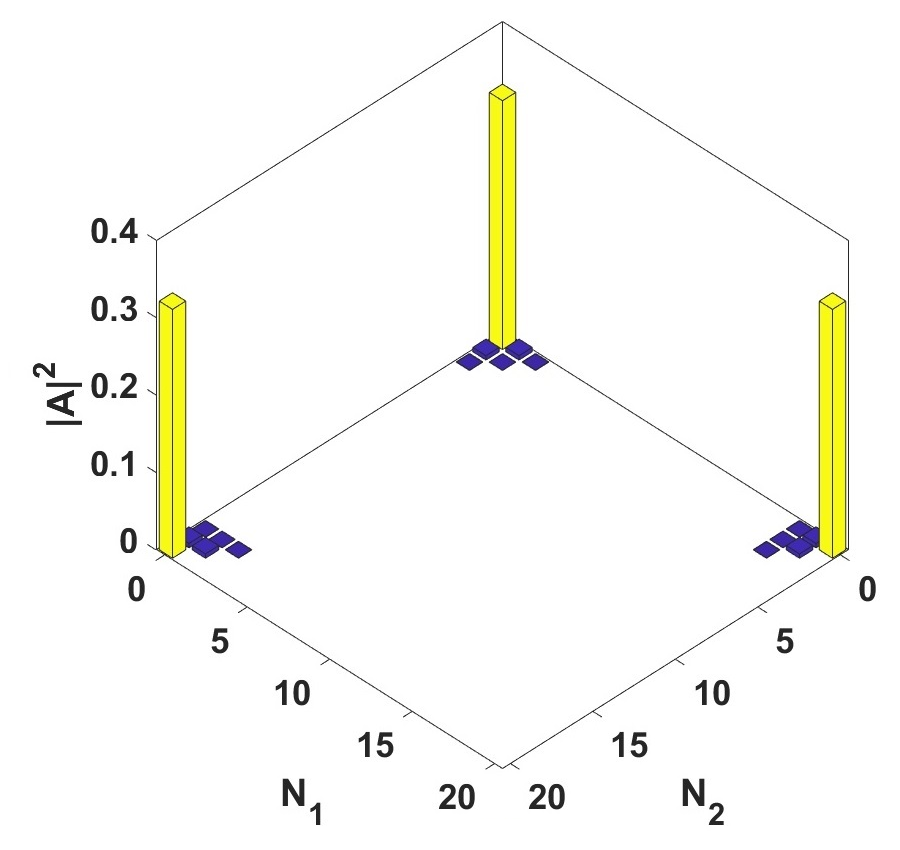}}\\ c) \end{minipage}
\caption{Distributions for TMSJJ ground state at (a) $\Lambda = 0$; (b) $\Lambda = \Lambda_{cr} \approx 3.30272$; (c) $\Lambda = 3.305$. $N=20$.}
\label{FIG:ground_states}
\end{figure*}

Coefficient $\alpha_{N_i,N_j}(\Lambda)$ corresponds to the energy of the intra-well particle interaction for the TMSJJ system with quantum numbers $N_1=N_i$, $N_2=N_j$, and $N_3=N-N_i-N_j$ at given $\Lambda$. The $\beta_{N_i,N_j}$ coefficient describes the inter-well interaction accompanied by a tunneling of a single particle from the $i$-th soliton to the $j$-th one. 

\begin{figure}[t]
\center{\includegraphics[width=1\linewidth]{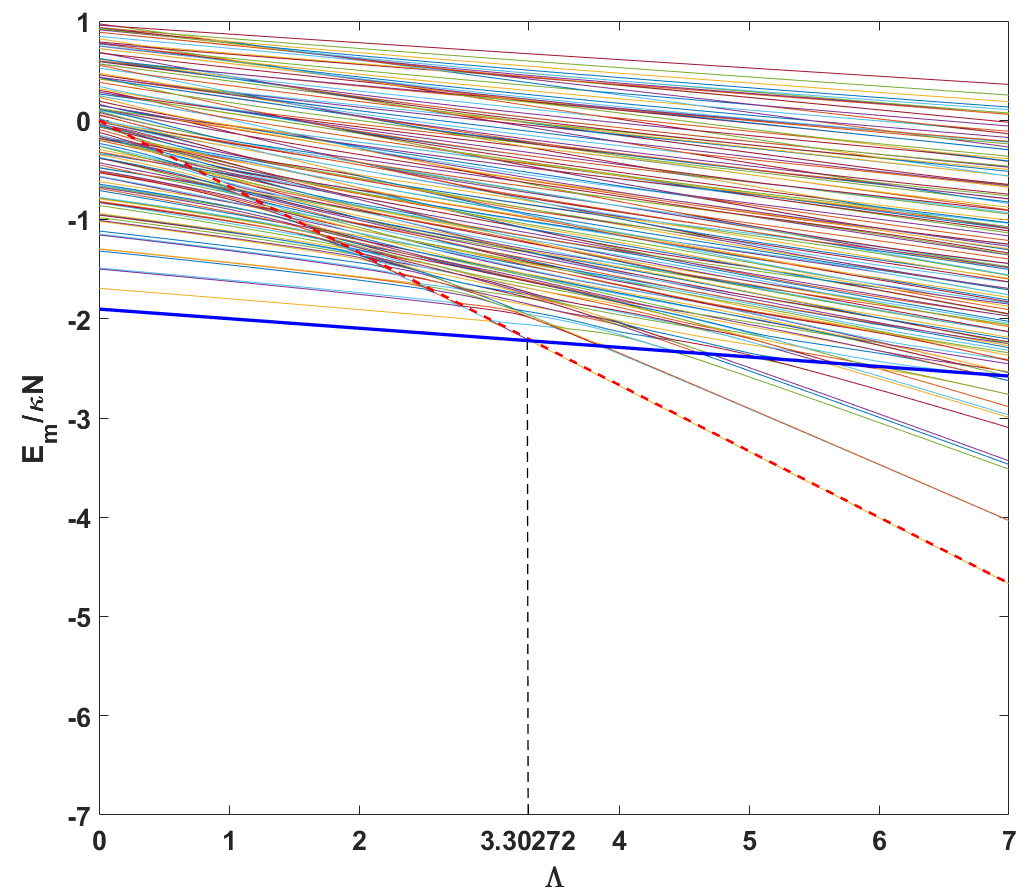}}
\caption{TMSJJ spectrum as a function of $\Lambda$; $N=20$. The phase-transition occurs at $\Lambda_{cr} = 3.30272$. The thick blue and dashed red lines denote the energies of atom-coherent and $N00N$ states}
\label{FIG:Spectrum}
\end{figure}

Hamiltonian~\eqref{Ham6} is then diagonalized, and one obtains $N$ energy eigenvalues $E_m$, represented in Fig.~\ref{FIG:Spectrum} as a function of tailoring parameter $\Lambda$. The ground state energy of the TMSJJ at $\Lambda<\Lambda_{cr}$ is $E/\kappa N \approx -1.911 - 0.008\Lambda$; it is marked by the solid blue line in Figure~\ref{FIG:Spectrum}. As seen from Fig.~\ref{FIG:ground_states}a, at $\Lambda<\Lambda_{cr}$ the TMSJJ ground state is Gaussian-like state that corresponds to the superfluid state of BECs; the particles tend to equally populate all three wells. In Fig.~\ref{FIG:Spectrum} the quantum phase transition is clearly seen to occur at $\Lambda = \Lambda_{cr}\approx 3.30272$, similarly to the one in the two-mode SJJ model at $\Lambda \approx 2.0009925$, cf.~\cite{Tsarev2020}. 

In Fig.~\ref{FIG:ground_states} we establish the ground state behaviour for the quantum TMSJJ system nearby critical point $\Lambda_{cr}$. In particular, at $\Lambda = \Lambda_{cr}$ the transition to three-mode entangled Fock state occurs; all $N$ particles tend to populate ``edges" $ $ $\left|N_1,0,0\right\rangle$, $\left|0,N_2,0\right\rangle$ and $\left|0,0,N_3\right\rangle$ in the Fock state basis. In this limit both Gaussian-like and $N00N$ states of the TMSJJ possess the same energy, and, thus, the resulting ground state represents a coherent superposition of them, see Fig.~\ref{FIG:Spectrum} and Fig.~\ref{FIG:ground_states}b. 
 
At $\Lambda>\Lambda_{cr}$ the $N00N$ state, that is 
\begin{equation}\label{n00n2}
\left|N00N\right\rangle = \frac{1}{\sqrt{3}}\left( \left|N,0,0\right\rangle + \left|0,N,0\right\rangle + \left|0,0,N\right\rangle\right),
\end{equation}
becomes energetically favorable for the ground state of solitons, see Fig.~\ref{FIG:ground_states}c. The $N00N$ state~\eqref{n00n2} possesses the energy 
\begin{equation}\label{energy}
 E = - \kappa N\frac{2\Lambda}{3}, 
\end{equation}
indicated with the lower part of red dashed line in Fig.~\ref{FIG:Spectrum}. Noteworthy, at $\Lambda\leq2.08$ the Hartree approach predicts $N00N$ state~\eqref{TMSJJ_n00n} that corresponds to some excited levels in Fig.~\ref{FIG:Spectrum}. Roughly speaking, the value of the $\Lambda$-parameter for these states corresponds to the same energy~\eqref{energy} for the upper part of the red dashed line in Fig.~\ref{FIG:Spectrum}. Thus, state~\eqref{n00n2} represents the tripartite $N00N$ state obtained in~\eqref{TMSJJ_n00n} at $\Theta_{\pm}=0$. We will use the $N00N$ state in~\eqref{n00n2} as a probe one, $\left|\psi_{in}\right\rangle$ (see~\eqref{psi_in}), for two-parameter metrological purposes, cf. Fig.~\ref{FIG:Scheme}a.

The feasibility of achieving $\Lambda_{cr}$ in current experiments with bright solitons is critical for this work. Notably, usual (two-mode) condensate Josephson junctions, which pose negative scattering length, enable to obtain the $N00N$ state in the limit of $\Lambda\gg1$, that implies a large number of particles, see e.g.~\cite{Tsarev2019}. Practically, this limit is hardly achievable with attractive condensate particles due to the condensate wave function collapse that occurs at $N\simeq 5\times 10^3$ for lithium condensates~\cite{Strecker2002,Khaykovich2002}. Contrary, the $N00N$ states based on matter-wave bright solitons may be observed with BEC solitons possessing mesoscopic number of particles (up to thousand), cf.~\cite{Alodjants2022,Kevrekidis2008,Strecker2002,Khaykovich2002,Nguyen2014}. The value of negative scattering length may be tailored by means of the Feshbach resonance technique, cf.~\cite{Grimm}. Thus, critical value $\Lambda_{cr}$ for the three soliton model may be obtained in the same manner as we previously discussed in~\cite{Tsarev2020} for the SJJ system.

Noteworthy, Fig.~\ref{FIG:ground_states} - Fig.~\ref{FIG:LQM}, that illustrate the main results in this work, are plotted for physically small number of particles $N=20$ because of lacking computational facilities. Formally, such a number of particles in real-world experiments require extremely large soliton nonlinearity, cf.~\cite{Kevrekidis2008}. However, the key physical features we discuss through this work for coupled solitons remain unchanged with $N$ increasing as a parameter. Thus, we expect the obtained results to be valid for mesoscopic number of particles (up to thousand) when bright solitons are stable and may be formed in condensate with negative scattering length, cf.~\cite{Khaykovich2002}.

\section{\label{sec:level1}Lossy quantum metrology with TMSJJ}

\begin{figure*}[t]
\begin{minipage}[h]{0.49\linewidth}
\center{\includegraphics[width=1\linewidth]{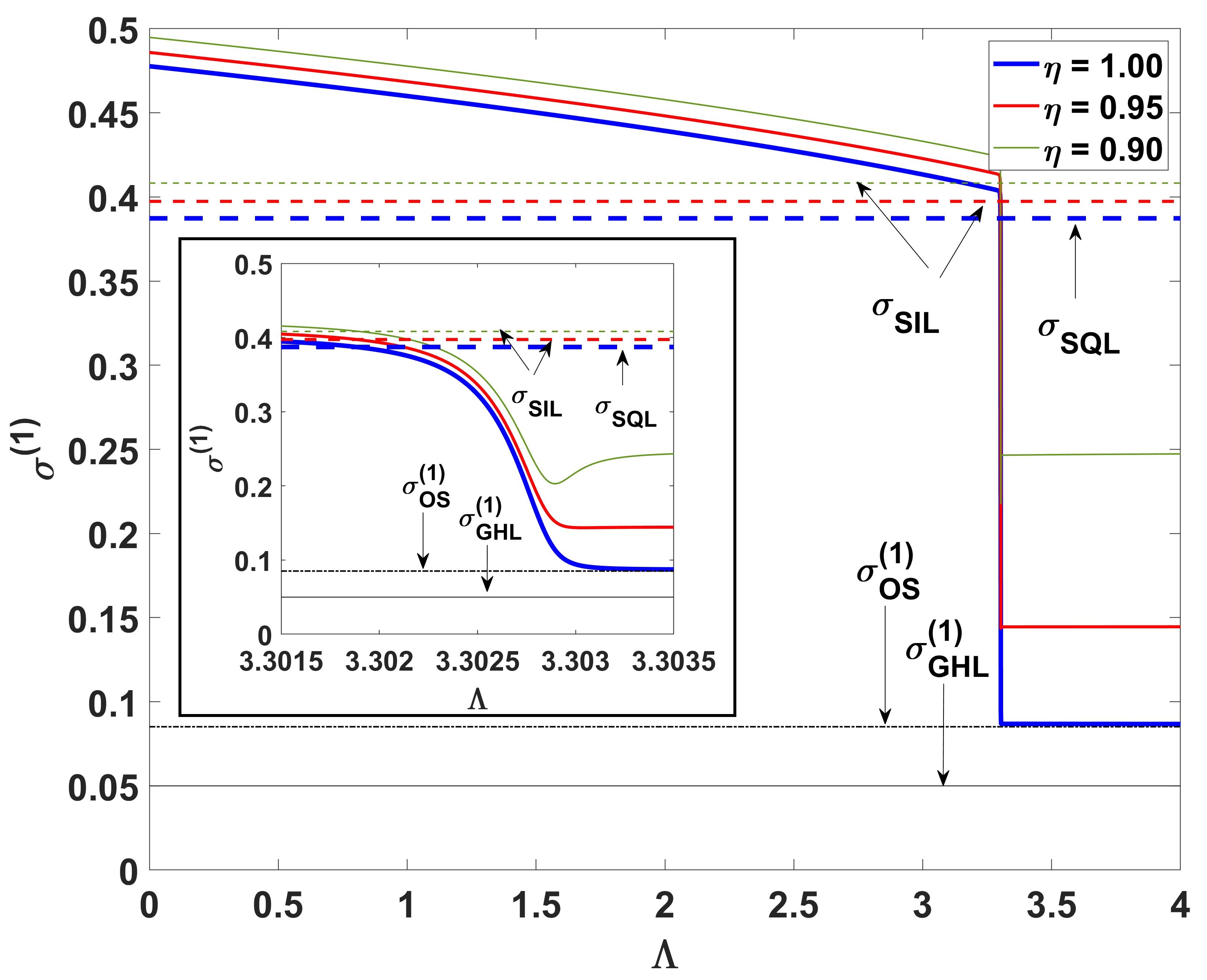}}\\ a) \end{minipage}
\hfill
\begin{minipage}[h]{0.49\linewidth}
\center{\includegraphics[width=1\linewidth]{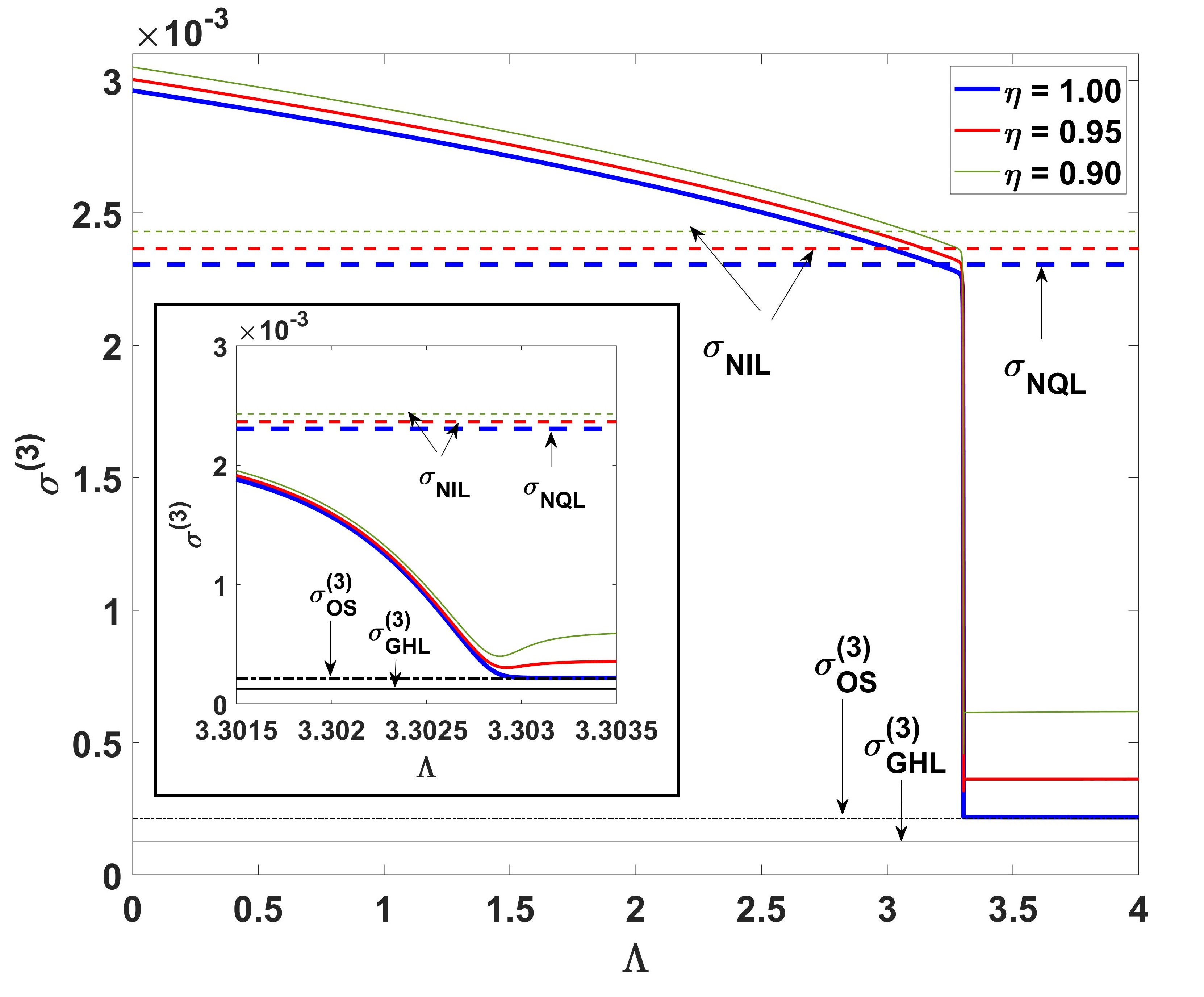}} \\b)
\end{minipage}
\caption{Accuracy bound $\sigma^{(k)}$ vs. vital parameter $\Lambda$ for (a) linear ($k=1$) and, (b) nonlinear ($k=3$) quantum metrology protocols with solitons, respectively. The losses are characterized by deviation of $\eta$-parameter from unity. Number of particles is $N=20$. Limiting linear quantum metrology characterizes by SQL ( $\sigma_{SQL} = \sqrt{3/N}$) and SIL ( $\sigma_{SIL} = \sqrt{3/\eta N}$ ), the dashed lines in (a). Nonlinear quantum metrology described by means of NQL ($\sigma_{NQL} \approx \sqrt{27/N^5}$) and NIL ($\sigma_{NIL} \approx \sqrt{27/\eta N^5}$), the dashed lines in (b). In both cases the black dashed-and-dotted lines denote optimized states accuracy $\sigma_{OS}^{(k)} = \sqrt{2.9}/N^k$ and thin solid black lines denote GHL $\sigma_{GHL}^{(k)} = 1/N^k$. The insets demonstrate accuracy bounds $\sigma^{(k)}$ in the vicinity of critical point $\Lambda=\Lambda_{cr}$. Other details are given in the text.}
\label{FIG:LQM}
\end{figure*}

Here, we examine a practically feasible two-parameter quantum metrology problem with solitons in the presence of losses. We do not consider the case when loses and decoherence occur under the probe (multipartite) state, $\left|\psi_{in}\right\rangle$, preparation for further metrological implementation, see Fig.~\ref{FIG:Scheme}a. We discussed in detail how atomic condensates are suitable for $\left|\psi_{in}\right\rangle$ preparation in the two-mode limit in~\cite{Tsarev2020}, see also \cite{Ngo2021, Ngo}. Below we assume that losses may appear in the scheme during probe state $\left|\psi_{in}\right\rangle$ evolution, see Fig.~\ref{FIG:Scheme}a. The metrology protocol consists of three steps. The first one corresponds to three-mode $N00N$ state $\left|\psi_{in}\right\rangle=\left|N00N\right\rangle$ preparation that may be realized by the TMSJJ device, see~\eqref{n00n2}. Then, we assume that two modes accumulate relative phases, that are $\chi$-dependent, while one (reference) mode remains unshifted, see Eq.~\eqref{psi_n}. Finally, the third step requires some linear operation $\hat{U}_L$ to mix all the modes together and make them interfere. For three modes ($n=3$) we can use a so-called tritter, which is familiar in quantum optics and may be designed by a photonic circuit~\cite{Spagnolo}. 

We exploit fictitious beam splitter (FBS) method for accounting particle losses in the scheme in Fig.~\ref{FIG:Scheme}a. Consider three FBSs, which impose equal transparency parameter $0<\eta\leq1$; the ideal lossless quantum metrology limit corresponds to value $\eta=1$. We assume that each FBS acts on a separate channel of the interferometer, transforming the corresponding Fock state as follows (cf.~\cite{Dobrzanski2009}):
\begin{equation}
 \left | m \right \rangle \rightarrow \sum_{l=0}^m \sqrt{\binom{m}{l} \eta^m \left( \eta^{-1}-1 \right)^l} \left | m-l\right \rangle \otimes \left | l \right \rangle,
\end{equation}
where $m$ is the initial population of the mode; $l$ is the number of particles lost; and $\binom{m}{l}= \frac{m!}{l!\left(m-l\right)!}$. Notice, in the optical experiment actual beam splitters can be used to model losses. In this case, the number of particles lost from each mode $l_i$, $i=1,2,3$, can be measured with photon number resolving detectors.

To be more specific, we consider lossy quantum metrology operating with the input state~\eqref{Fock-vector} generated by the TMSJJ. We consider the third mode as the reference one (c.f. Fig.\ref{FIG:Scheme}a). Other two modes accumulate a relative phase shift described by operator
\begin{equation}\label{PS}
 \hat{U}_{PS} = \exp\left[i\chi_1\left(\hat{a}_1^\dag\hat{a}_1\right)^k + i\chi_2\left(\hat{a}_2^\dag\hat{a}_2\right)^k\right].
\end{equation}

Noteworthy, operator~\eqref{PS} commutes with the Kraus operator that describes the particle losses within FBS approach. Therefore, it is not important where particles are lost; it may happen before or after the phase accumulation, cf.~\cite{Dobrzanski2009}.

Since we are not interested in the lost particles, we can trace them out and consider the mixed output quantum state with density matrix
\begin{subequations}\label{denc_matrix}
\begin{equation}
\rho=\sum_{l_1=0}^N\sum_{l_2=0}^{N-l_1}\sum_{l_3=0}^{N-l_1-l_2} p_{l_1,l_2,l_3}\left | \xi_{l_1,l_2,l_3} \right \rangle \left \langle \xi_{l_1,l_2,l_3} \right |;
\end{equation}
\begin{widetext}
\begin{equation}
\left | \xi_{l_1,l_2,l_3} \right \rangle = \frac{1}{\sqrt{p_{l_1,l_2,l_3}}} \sum_{N_1=l_1}^{N-l_2-l_3}\sum_{N_2=l_2}^{N-N_1-l_3} A_{N_1,N_2}\sqrt{B_{l_1,l_2,l_3}^{N_1,N_2}}e^{i\chi_1 N_1^k + i\chi_2 N_2^k}\left |N_1-l_1,N_2-l_2,N_3-l_3 \right \rangle,
\end{equation}
\end{widetext}
\end{subequations}
where $N_3 = N - N_1 - N_2$, 
\begin{equation}
B_{l_1,l_2,l_3}^{N_1,N_2}=\binom{N_1}{l_1}\binom{N_2}{l_2}\binom{N_3}{l_3} \eta^N \left(\eta^{-1}-1 \right)^{l}
\end{equation}
with $l = l_1+l_2+l_3$; 
\begin{center}
$p_{l_1,l_2,l_3}=\sum_{N_1=l_1}^{N-l_2-l_3}\sum_{N_2=l_2}^{N-N_1-l_3} \left | A_{N_1,N_2} \right |^2 B_{l_1,l_2,l_3}^{N_1,N_2} $
\end{center}
is the probability to lose exactly $l_1$, $l_2$, and $l_3$ particles from the three interferometer channels.

For the QFI, we restrict ourselves only by its upper bound, $\tilde{F}$, that looks like 
\begin{widetext}
\begin{align}\label{QFI_UB}
F_{ij}\leq \widetilde{F}_{ij} = 4 \sum_{l_1=0}^N\sum_{l_2=0}^{N-l_1}\sum_{l_3=0}^{N-l_1-l_2} p_{l_1,l_2,l_3}\big[\left\langle \partial_{\chi_i}\xi_{l_1,l_2,l_3} | \partial_{\chi_j}\xi_{l_1,l_2,l_3}\right\rangle -\left\langle\partial_{\chi_i}\xi_{l_1,l_2,l_3}|\xi_{l_1,l_2,l_3}\right\rangle\left\langle\xi_{l_1,l_2,l_3}|\partial_{\chi_j}\xi_{l_1,l_2,l_3}\right\rangle\big].
\end{align}
Substituting~\eqref{denc_matrix} into~\eqref{QFI_UB} we obtain
\begin{eqnarray}\label{QFI_UB2}
\widetilde{F}_{ij}&=&4\sum_{N_1=0}^N \sum_{N_2=0}^{N-N_1} (N_iN_j)^k \left | A_{N_1,N_2} \right |^2\\
&-&4\sum_{l_1=0}^N \sum_{l_2=0}^{N-l_1} \sum_{l_3=0}^{N-l_1-l_2} \frac{\left(\sum_{N_1=l_1}^{N-l_2-l_3}\sum_{N_2=l_2}^{N-N_1-l_3}N_i\left|A_{N_1,N_2}\right|^2B_{l_1,l_2,l_3}^{N_1,N_2}\right)\left(\sum_{N_1=l_1}^{N-l_2-l_3}\sum_{N_2=l_2}^{N-N_1-l_3}N_j\left|A_{N_1,N_2}\right|^2B_{l_1,l_2,l_3}^{N_1,N_2}\right)}{\sum_{N_1=l_1}^{N-l_2-l_3} \sum_{N_2=l_2}^{N-N_1-l_3} \left | A_{N_1,N_2} \right |^2 B_{l_1,l_2,l_3}^{N_1,N_2}}. \nonumber
\end{eqnarray}
\end{widetext}
Notice, at $\eta = 1$ (i.e. without particle losses) $B_{0,0,0}^{N_1,N_2} = 1$ and $B_{l_1,l_2,l_3}^{N_1,N_2} = 0$ for any $l_{1,2,3} > 0$. In this case $\tilde{F} = F$. 

Coefficients $A_{N_1,N_2}$ can be obtained by numerical simulation of Eqs.~\eqref{SC_eq2} and~\eqref{SC_eq3} for various $\Lambda$. Fig.~\ref{FIG:LQM} exhibits the principal results of this work. It demonstrates the capability of the TMSJJ for quantum state preparation,which is relevant to quantum metrology with solitons. In particular, Fig.~\ref{FIG:LQM}a characterizes the linear quantum metrology, while Fig.~\ref{FIG:LQM}b describes the nonlinear quantum metrology approach. We represent the accuracy bound for the $\chi$-parameter measurement and estimation as a function of $\Lambda$ for different $\eta$:
\begin{equation}
\sigma^{(k)} = \left(\textrm{Tr}\left(\hat{\widetilde{F}}^{-1}\right)\right)^{1/2},
\end{equation}
where $\hat{\widetilde{F}} \equiv \{\widetilde{F}_{ij}\}$ is the QFI upper bound matrix.

The thick blue curves in Fig.~\ref{FIG:LQM} are relevant to $\eta = 1$ limit, characterizing the maximal metrological capacity that may be achieved without losses in general. In particular, the upper thick blue dashed line characterizes the SQL within the linear metrology approach, and nonlinear SQL for the nonlinear one. Both of them may be attained with coherent states, cf.~\cite{Humphreys}. One can estimate these limits in the case of two-parameter metrology based on Gaussian three-mode quantum state
\begin{widetext}
\begin{eqnarray}
\left|\psi\right\rangle = \sum_{N_1=0}^N\sum_{N_2=0}^{N-N_1}\sqrt{p(N_1,N_2)} e^{i\chi_1 N_1^k + i\chi_2 N_2^k}\big|N_1,N_2,&&N-N_1-N_2\big\rangle,
\end{eqnarray}
where
\begin{eqnarray}\label{Gauss3}
p(N_1,N_2) = \frac{9}{2\sqrt{3}\pi N}\exp\Bigg[-\frac{9}{4N}\left(N_1+N_2-\frac{2N}{3}\right)^2 - \frac{3}{4N}\left(N_1-N_2\right)^2\Bigg]
\end{eqnarray}
\end{widetext}
characterizes Gaussian distribution function for $N\gg1$. Eq.~\eqref{Gauss3} implies $\sigma_{SQL} = \sqrt{3/N}$ for the linear quantum metrology ($k=1$) approach, and $\sigma_{NQL} \approx \sqrt{27/N^5}$ for the nonlinear one, $k=3$. 

In the presence of weak losses $\sigma_{SQL}$ and $\sigma_{NQL}$ establish standard (SIL) and nonlinear (NIL) interferometric limits, respectively; they are 
\begin{equation}\label{SIL1}
\sigma_{SIL} = \sqrt{\frac{3}{\eta N}};
\end{equation}
\begin{equation}\label{NIL3}
\sigma_{NIL} \approx \sqrt{\frac{27}{\eta N^5}}.
\end{equation}
 
Also in Fig.~\ref{FIG:LQM} we focus on the area nearby the critical value $\Lambda_{cr}\approx 3.30272$ that corresponds to the phase transition to the $N00N$ state for the TMSJJ; this area is zoomed in the insets to Fig.~\ref{FIG:LQM}a and Fig.~\ref{FIG:LQM}b, respectively. 

In general, without of losses accuracy $\sigma^{(k)}$ approaches the optimal state level (see the thick blue curves in Fig.~\ref{FIG:LQM}). Fig.~\ref{FIG:LQM} clearly demonstrates that accuracy $\sigma^{(k)}$ beats vital classical interferometric limits~\eqref{SIL1},~\eqref{NIL3} for $\Lambda > \Lambda_{cr}$ even in the presence of moderate losses, i.e. when an almost $N00N$ state is prepared by the TMSJJ. 

Finally, let us examine the measurement and estimation procedure with states $\left|N00N\right\rangle_{\pm}$ capable for $\left|\psi_n\right\rangle$ formation as a result, see~\eqref{psi_n},~\eqref{TMSJJ_n00n}. We assume that the three-mode $N00N$ state possesses phase-shifts $\Theta_\pm$ providing a unique opportunity to estimate parameter $\chi \equiv \Lambda/N^2 \equiv u^2/16\kappa$. Since there is only one $\chi$ parameter to be measured effectively, the QFIs may be calculated as
\begin{eqnarray}
\nonumber
&&F_\pm = 4\Big[{}_\pm\left\langle\partial_\chi N00N|\partial_\chi N00N\right\rangle{}_\pm \\
&&\;\;\;\;\;\;\;\;\;\;\;\;\;\;\;\;\;\;\;\;\;\;\;\;\;\; - {\big|{}_\pm\langle\partial_\chi N00N| N00N\rangle{}_\pm\big|^2}\Big],
\end{eqnarray}
where $F_+$ and $F_-$ are the QFIs for in- and out-of-phase solitons, respectively; $\left|\partial\chi N00N\right\rangle{}_\pm \equiv \frac{\partial}{\partial\chi}\left|N00N\right\rangle{}_\pm=\frac{\partial \Theta_\pm}{\partial\chi}\frac{\partial}{\partial\Theta_\pm}\left|N00N\right\rangle{}_\pm$; $\left|N00N\right\rangle{}_\pm$ is state~\eqref{TMSJJ_n00n}, and phases $\Theta_\pm$ obey~\eqref{out_of_phase-n00n} and~\eqref{in_phase-n00n}. 

After some straightforward calculations for the accuracies of the $\chi$ parameter measurement and estimation using the in-phase ($\sigma_{\chi_+}$) and out-of-phase ($\sigma_{\chi_-}$) solitons configurations, we obtain
\begin{equation}\label{pm1}
\sigma_{\chi_+} = 1.58/N^3,
\end{equation}
\begin{equation}\label{pm}
\sigma_{\chi_-} = 1.22/N^3,
\end{equation}
respectively. Remarkably, for both cases accuracy is inversely proportional to $N^3$ that represents the metrological limit of phase estimation for interacting solitons, cf. Fig.~\ref{FIG:LQM}a and Fig.~\ref{FIG:LQM}b,~\cite{Tsarev2018}. Notably, some improvement of accuracy $\sigma_{\chi_-}$ (in comparison with $\sigma_{\chi_+}$) appears due to the nonlinear soliton phase counter-accumulation, cf.~\cite{Jarzyna}.

\section{\label{sec:level1}Conclusion}

In this work, we have considered the $d$-parameter quantum metrology problem ($d>1$) with sensor networks operating with bright solitons. The GHL is introduced for both linear and nonlinear quantum metrology tasks. In this framework, we have first examined the multipartite $N00N$ state distributed over QSN. Notably, general strategies, which use multipartite $N00N$ states, demonstrate the $\sqrt{\frac{d(d+1)}{2}}$ times accuracy degradation in the $d$ parameters measurement and estimation problem, see~\eqref{QFI3}. Thus, we have shown that the balanced $N00N$ state is not optimal even without losses in this case. However, for the QSNs, which use coupled solitons, with moderate $d$, the accuracy is close to the fundamental GHL established in this work. To be more specific, we have considered the three-mode soliton Josephson junction (TMSJJ) model that allows preparation the tripartite $N00N$-like (probe) state suitable for the two parameter metrology problem. The TMSJJ represents a generalization of the two-mode soliton Josephson junction system established for three weakly coupled solitons, cf.~\cite{Alodjants2022}. In quantum optics such a model is valid for solitons propagating in coupled optical fibers or waveguides. We have shown that the TMSJJ exhibits the quantum phase transition to the superposition of entangled Fock states capable for the three-mode $N00N$ state formation with mesoscopic number of particles. The phase transition occurs at some critical value $\Lambda_{cr}$ of dimensionless parameter $\Lambda$ that may be obtained within the current experiments with weakly coupled atomic condensates or optical beams in highly nonlinear Kerr-like materials. We have shown that beyond the critical value of parameter $\Lambda$ accuracy $\sigma^{(k)}$ approaches  the optimal state even in the presence of weak losses. We have also provided the quantum metrology protocol of the Kerr-like nonlinear $\chi$-parameter measurement and estimation within the in-phase and out-of-phase soliton configurations. It is shown that the best accuracy of the measurement is close to the GHL (see \eqref{pm}), which we can achieve with the out-of-phase interacting solitons. Our findings open new prospects for the problems of spatially distributed quantum sensing and metrology.

\section*{Acknowledgements}

Russian authors acknowledge support of Secs. III-IV from the Ministry of Science and Higher Education of the Russian Federation and South Ural State University (Agreement No. 075-15-2022-1116), and Sec. II from Project 075-15-2021-1349.


\begin{thebibliography}{54}

\bibitem{PezzeRMP} 
L. Pezz{\`e}, A. Smerzi, M. K. Oberthaler \textit{et al.}, Quantum metrology with nonclassical states of atomic ensembles, Rev. Mod. Phys. {\bf 90}, 035005 (2018).

\bibitem{DegenRMP}
C. L. Degen, F. Reinhard, and P. Cappellaro, Quantum sensing, Rev. Mod. Phys. {\bf 89}, 035002 (2017).

\bibitem{Crawford}
S. E. Crawford, R. A Shugayev, H. P. Paudel \textit{et al.}, Quantum Sensing for Energy Applications: Review and Perspective. Adv. Quantum Technol., {\bf 4}, 2100049 (2021).

\bibitem{Bongs}
K. Bongs, M. Holynski, J. Vovrosh \textit{et al.}, Taking atom interferometric quantum sensors from the laboratory to real-world applications. Nature Rev. Phys. {\bf 1}, 731–739 (2019).

\bibitem{Wehner}
S. Wehner, D. Elkouss, and R. Hanson, Quantum internet: A vision for the road ahead, Science, {\bf 362}, 6412 (2018).

\bibitem{Becker}
D. Becker, M.D. Lachmann, S.T. Seidel \textit{et al.}, Space-borne Bose–Einstein condensation for precision interferometry. Nature {\bf562}, 391 (2018).

\bibitem{Polino}
E. Polino, M. Valeri, N. Spagnolo, and F. Sciarrino, Photonic quantum metrology, AVS Quantum Science {\bf 2}, 024703 (2020).

\bibitem{Dowling}
J. P. Dowling, Quantum optical metrology–the lowdown on high-N00N states, Contemporary physics {\bf 49}, 125-143 (2008).

\bibitem{Gerry2021}
Richard J. Birrittella, Paul M. Alsing, and Christopher C. Gerry, The parity operator: Applications in quantum metrology, AVS Quantum Sci. {\bf 3}, 014701 (2021).

\bibitem{Tsarev2018}
D. V. Tsarev, S. M. Arakelian, You-Lin Chuang, Ray-Kuang Lee, and A. P. Alodjants, Quantum metrology beyond Heisenberg limit with entangled matter wave solitons, Opt. Express {\bf 26}, 19583 (2018).

\bibitem{Ngo2021}
T.V. Ngo, D.V. Tsarev, Ray-Kuang Lee, A.P. Alodjants. Bose–Einstein condensate soliton qubit states for metrological applications. Sci. Rep., {\bf 11}, 19363 (2021).

\bibitem{Grassl}
A. J. Scott, M. Grassl, Symmetric informationally complete positive-operator-valued measures: A new computer study J. Math. Phys. {\bf 51}, 042203 (2010).

\bibitem{Planat} 
M. Planat, Z. Gedik, Magic informationally complete POVMs with permutations.R. Soc. open sci. {\bf 4}, 170387 (2017).

\bibitem{Yoshida}
Masakazu Yoshida and Gen Kimura, Construction of general symmetric-informationally-complete–positive-operator-valued measures by using a complete orthogonal basis, Phys. Rev. A {\bf 106}, 022408 (2022).

\bibitem{Pinto}
Douglas F. Pinto, Marcelo S. Zanetti, Marcos L. W. Basso, and Jonas Maziero, Simulation of positive operator-valued measures and quantum instruments via quantum state-preparation algorithms, Phys. Rev. A {\bf 107}, 022411 (2023).

\bibitem{Zhu}
Huangjun Zhu, SIC POVMs and Clifford groups in prime dimensions, J. Phys. A: Math. Theor. {\bf 43}, 305305 (2010).

\bibitem{Renes}
Joseph M. Renes, Robin Blume-Kohout, A. J. Scott, and C. M. Caves, Symmetric informationally complete quantum measurements, Journal of Mathematical Physics {\bf 45}, 2171 (2004).

\bibitem{Czerwinski}
Artur Czerwinski, Quantum state tomography with informationally complete POVMs generated in the time domain. Quantum Information Processing, {\bf 20}, 105 (2021).

\bibitem{Ferrie} 
Christopher Ferrie and Robin Blume-Kohout, Minimax Quantum Tomography: Estimators and Relative Entropy Bounds. Phys. Rev. Lett. {\bf 116}, 090407 (2016).

\bibitem{Brida} 
Yu. I. Bogdanov, G. Brida, M. Genovese, S. P. Kulik, E. V. Moreva, and A. P. Shurupov, Statistical estimation of the efficiency of quantum state tomography protocols. Phys. Rev. Lett.,{\bf 105}, 010404 (2010).

\bibitem{Lundeen}
J. Lundeen, A. Feito, H. Coldenstrodt-Ronge, et al. Tomography of quantum detectors. Nature Phys {\bf 5}, 27 (2009).

\bibitem{Leuchs}
N. Bent, H. Qassim, A.A. Tahir, D. Sych, G. Leuchs, L.L. Sánchez-Soto, E. Karimi, and R.W. Boyd, Experimental Realization of Quantum Tomography of Photonic Qudits via Symmetric Informationally Complete Positive Operator-Valued Measures, Phys. Rev. X {\bf 5}, 041006 (2015).

\bibitem{Tabia} 
Gelo Noel M. Tabia, Experimental scheme for qubit and qutrit symmetric informationally complete positive operator-valued measurements using multiport devices, Phys. Rev. A {\bf 86}, 062107 (2012).

\bibitem{Padua}
W. M. Pimenta, B. Marques, T. O. Maciel, R. O. Vianna, A. Delgado, C. Saavedra, and S. Pádua, Minimum tomography of two entangled qutrits using local measurements of one-qutrit symmetric informationally complete positive operator-valued measure, Phys. Rev. A {\bf 88}, 012112 (2013).

\bibitem{Steinberg}
Z. E. D. Medendorp, F. A. Torres-Ruiz, L. K. Shalm, G. N. M. Tabia, C. A. Fuchs, and A. M. Steinberg, Experimental characterization of qutrits using symmetric informationally complete positive operator-valued measurements, Phys. Rev. A {\bf 83}, 051801(R) (2011).

\bibitem{Ling}
Alexander Ling, Kee Pang Soh, Antía Lamas-Linares, and Christian Kurtsiefer, Experimental polarization state tomography using optimal polarimeters, Phys. Rev. A {\bf 74}, 022309 (2006).

\bibitem{Okamoto}
R. Okamoto, H. F. Hofmann, T. Nagata \textit{et al.}, Beating the standard quantum limit: phase super-sensitivity of N-photon interferometers, New J. Phys. {\bf 10}, 073033 (2008).

\bibitem{Maldonado}
D. Maldonado-Mundo and A. Luis, Metrological resolution and minimum uncertainty states in linear and nonlinear signal detection schemes, Phys. Rev. A {\bf 80}, 063811 (2009).

\bibitem{Napolitano} 
M. Napolitano and M. W. Mitchell, Nonlinear metrology with a quantum interface, New J. Phys. {\bf 12}, 09301 (2010). 

\bibitem{Tsarev2019}
D. V. Tsarev, T. V. Ngo, R. K. Lee, and A. P. Alodjants, Nonlinear quantum metrology with moving matter-wave solitons, New J. Phys. \textbf{21} 083041 (2019).

\bibitem{Alodjants2022}
A. P. Alodjants, D. V. Tsarev, T. V. Ngo, and R. K. Lee, Enhanced nonlinear quantum metrology with weakly coupled solitons in the presence of particle losses, Phys. Rev. A {\bf 105}, 012606 (2022).

\bibitem{Mazzarella}
G. Mazzarella, L. Salasnich, A. Parola, and F. Toigo, Coherence and entanglement in the ground state of a bosonic Josephson junction: From macroscopic Schr{\"o}dinger cat states to separable Fock states, Phys. Rev. A {\bf 83}, 053607 (2011). 

 \bibitem{Drummond}
Q. Y. He, M. D. Reid, T. G. Vaughan \textit{et al.}, Einstein-Podolsky-Rosen Entanglement Strategies in Two-Well Bose-Einstein Condensates, Phys. Rev. Lett. {\bf 106}, 120405 (2011).

\bibitem{Zheshen}
Z. Zhang and Q. Zhuang, Distributed Quantum Sensing, Quantum Science Technology {\bf 6}, 043001 (2021). 

\bibitem{Goldberg}
Aaron Z. Goldberg, I. Gianani, M. Barbieri, F. Sciarrino, A.M. Steinberg, and N. Spagnolo, Multiphase estimation without a reference mode, Phys. Rev. A {\bf 102}, 022230  (2020).

\bibitem{Amico}
L. Amico, M. Boshier, G. Birkl \textit{et al.}, Roadmap on Atomtronics: State of the art and perspective featured, AVS Quantum Science {\bf 3}, 039201 (2021).

\bibitem{Flamini}
F. Flamini, N. Spagnolo, and F. Sciarrino, Photonic quantum information processing: a review, Reports on Progress in Physics, {\bf 82}, 016001 (2019).

\bibitem{Clements}
W. R. Clements, P. C. Humphreys, B. J. Metcalf \textit{et al.}, Optimal design for universal multiport interferometers, Optica {\bf 3}, 1460-1465 (2016).

\bibitem{Fldzhyan}
S. A. Fldzhyan, M. Yu. Saygin, and S. P. Kulik, Optimal design of error-tolerant reprogrammable multiport interferometers, Opt. Lett. {\bf 45}, 2632-2635 (2020).

\bibitem{Moss}
E. B. Corcoran, M. Tan, X. Xu \textit{et al.}, Ultra-dense optical data transmission over standard fibre with a single chip source, Nat Commun. {\bf 11}, 2568 (2020).

\bibitem{Karpov}
M. Karpov M.H.P. Pfeiffer, H. Guo, et al. Dynamics of soliton crystals in optical microresonators. Nat. Phys. {\bf 15}, 1071 (2019).

\bibitem{Lukin}
M. A. Guidry, D. M. Lukin, K. Y. Yang \textit{et al.}, Quantum optics of soliton microcombs, Nat. Photon. {\bf 16}, 52 (2022).

\bibitem{Jing}
J. Liu, X. M. Lu, Z. Sun, and X. Wang, Quantum multiparameter metrology with generalized entangled coherent state, J. Phys. A: Math. Theor. {\bf 49}, 115302 (2016).

\bibitem{Gessner} 
M. Gessner, L. Pezz{\`e}, and A. Smerzi, Sensitivity Bounds for Multiparameter Quantum Metrology, Phys. Rev. Lett. {\bf 121}, 130503 (2018).

\bibitem{Pezze}
L. Pezz{\`e}, Entanglement-enhanced sensor networks, Nat. Photonics {\bf 15}, 74–76 (2021).

\bibitem{Bringewatt} 
J. Bringewatt, I. Boettcher, P. Niroula, P. Bienias, and A. V. Gorshkov, Protocols for estimating multiple functions with quantum sensor networks: Geometry and performance, Phys. Rev. Research {\bf 3}, 033011 (2021).

 \bibitem{Humphreys}
P. C. Humphreys, M. Barbieri, A. Datta, and I. A. Walmsley, Quantum Enhanced Multiple Phase Estimation, Phys. Rev. Lett. {\bf 111}, 070403 (2013).

\bibitem{Hong}
S. Hong, J. Ur. Rehman, Y.S. Kim \textit{et al.}, Quantum enhanced multiple-phase estimation with multi-mode N00N states, Nat. Commun. {\bf 12}, 5211 (2021).

\bibitem{Tsarev2020} 
D. V. Tsarev, A. P. Alodjants, T. V. Ngo, and R. K. Lee, Mesoscopic quantum superposition states of weakly-coupled matter-wave solitons, New J. Phys. \textbf{22}, 113016 (2020).

\bibitem{Ngo} 
Vinh T. Ngo, D.V. Tsarev, A.P. Alodjants. Coupled Solitons for Quantum Communication and Metrology in the Presence of Particle Dissipation. J Russ. Laser. Res. {\bf 42}, 523 (2021).

\bibitem{Khaykovich2002}
L. Khaykovich, F. Schreck, G. Ferrari \textit{et al.}, Formation of a Matter-Wave Bright Soliton, Science {\bf 296}, 1290 (2002).

\bibitem{Spagnolo}
N. Spagnolo, L. Aparo, C. Vitelli \textit{et al.}, Quantum interferometry with three-dimensional geometry, Sci. Rep. 2, {\bf 862} (2012).

\bibitem{Compagno}
E. Compagno, G. Quesnel, A. Minguzzi, L. Amico and D. Feinberg, Multimode N00N states in driven atomtronic circuits, Phys. Rev. Research {\bf 2}, 043118 (2020).

\bibitem{Nemoto}
K. Nemoto, C.A. Holmes, G.J. Milburn, and W.J. Munro, Quantum dynamics of three coupled atomic Bose-Einstein condensates, Phys Rev. A, {\bf 63}, 013604 (2000).

\bibitem{Guo}
Q. Guo, X.Z. Chen, and B. Wu, Tunneling dynamics and band structures of three weakly coupled Bose-Einstein condensates, Optics Express {\bf 22}, 19234 (2014).

\bibitem{Sich}
M. Sich, D. N. Krizhanovskii, M. S. Skolnick \textit{et al.}, Observation of bright polariton solitons in a semiconductor microcavity. Nature Photon {\bf 6}, 50–55 (2012).

\bibitem{Skryabin}
N. N. Skryabin, I. V. Dyakonov, M. Yu. Saygin, and S. P. Kulik, Waveguide-lattice-based architecture for multichannel optical transformations, Opt. Express {\bf 29}, 26058-26067 (2021).

\bibitem{Liao}
K. Liao, X. Hu, T. Gan \textit{et al.}, Photonic molecule quantum optics, Adv. Opt. Photon. \textbf{12}, 60 (2020).

\bibitem{Raghavan2000} 
S. Raghavan and G. P. Agrawan, Switching and self-trapping dynamics of Bose-Einstein solitons, Journal of Modern Optics, {\bf 47}, 1155-1169 (2000).

\bibitem{Lai1989} 
Y. Lai and H. A. Haus, Quantum theory of solitons in optical fibers. I. Time-dependent Hartree approximation, Phys. Rev. A {\bf 40}, 844 (1989). 

\bibitem{Lai1989a} 
Y. Lai and H. A. Haus, Quantum theory of solitons in optical fibers. II. Exact solution, Phys. Rev. A 40, {\bf 854} (1989).

\bibitem{Alodjants1995} 
A. P. Alodjants and S. M. Arakelian, Quantum chaos and its observation in coupled optical solitons, Zh. Eksp. i Teor. Fiz., {\bf 107}, 1792-1825 (1995).

\bibitem{Paraoanu} 
G. S. Paraoanu, S. Kohler, F. Sols, and A. J. Leggett, The Josephson plasmon as a Bogoliubov quasiparticle, Journal of Physics B: Atomic, {\bf 34} 4689 (2001).

\bibitem{Strecker2002}
K. E. Strecker, G. B. Partridge, A. G. Truscott, and R. G. Hulet, Formation and propagation of matter-wave soliton trains, Nature {\bf 417}, 150 (2002).

\bibitem{Kevrekidis2008}
P. G. Kevrekidis, D. J. Frantzeskakis, and R. Carretero-Gonz{\'a}lez, \textit{Emergent Nonlinear Phenomena in Bose–Einstein Condensates,} (Springer-Verlag, 2008).

\bibitem{Nguyen2014}
J. H. Nguyen, P. Dyke, D. Luo \textit{et al.}, Collisions of matter-wave solitons, Nature Phys. {\bf 10}, 918 (2014). 

\bibitem{Grimm}
C. Chin, R. Grimm, P. Julienne, and E. Tiesinga, Feshbach resonances in ultracold gases, Rev. Mod. Phys. {\bf 82}, 1225 (2010).

\bibitem{Dobrzanski2009} 
R. Demkowicz-Dobrzanski, U. Dorner, B. J. Smith \textit{et al.}, Quantum phase estimation with lossy interferometers, Phys. Rev. A \textbf{80}, 013825 (2009).

\bibitem{Jarzyna}
M. Jarzyna and R. Demkowicz-Dobrzanski, Quantum interferometry with and without an external phase reference, Phys. Rev. A {\bf 85}, 011801(R) (2012).

\end{thebibliography}
\end{document}